\documentclass[nofootinbib]{revtex4}
\usepackage[dvips]{color}
\usepackage[dvips]{graphicx}
\usepackage{epsfig}
\usepackage{subfig}

\begin{document}

\title{Distance-Redshift Relations in an Anisotropic Cosmological Model}

\author{R. S. Menezes Jr.$^{1}$\footnote{rsmjr@ifba.edu.br}, C. Pigozzo$^2$\footnote{cpigozzo@ufba.br} and S. Carneiro$^{2}$\footnote{saulo.carneiro.ufba@gmail.com}}

\affiliation{$^{1}$Instituto Federal da Bahia, Salvador, BA, Brazil\\ $^{2}$Instituto de  F\'{\i}sica, Universidade Federal da Bahia, Salvador, Bahia, Brazil}

\date{\today}

\begin{abstract}
\noindent
In this paper we study an anisotropic model generated from a particular Bianchi type-III metric, which is a generalization of G\"odel's metric and an exact solution of Einstein's field equations. We analyse type Ia supernova data, namely the SDSS sample calibrated with the MLCS2k2 fitter, and we verify in which ranges of distances and redshifts the anisotropy could be observed. We also consider, in a joint analysis, the position of the first peak in the CMB anisotropy spectrum, as well as current observational constraints on the Hubble constant. We conclude that a small anisotropy is permitted by the data, and that more accurate measurements of supernova distances above z = 2 might indicate the existence of such anisotropy in the universe.
\end{abstract}

\maketitle

\section{Introduction}

The search for evidence of possible anisotropies in the universe has been studied by several authors \cite{ Nodland:1997cc,Birch,antoniou,koivisto,Land:2005ad,Kalus:2012zu,Cai:2011xs}. The importance of these studies lies in the fact that the discovery of a possible anisotropy could lead to a better
understanding of the mechanism of the Big Bang and of the small temperature fluctuations detected in the cosmic
microwave background radiation (CMB) \cite{Larson:2010gs}. In addition, such a discovery would bring down the belief in the validity of the cosmological principle that is the basis of many intepretations of observations.

In order to detect possible signatures of anisotropy, in this paper we consider an anisotropic model constructed from an anisotropic (but homogeneous) metric with expansion, which is an exact solution of Einstein's equations. This metric, whose first particular case was proposed by G\"{o}del in 1949 \cite{godel}, is classified as Bianchi III and was studied in a more general form by M. Rebou\c{c}as and J. Tiomno \cite{reboucas}, and by V. Korotkii and Y. Obukhov \cite{korotkii} (from now on we will call it the RTKO metric). In general it has non-zero rotation (but zero shear), as well as a conformal expansion. Here we will take the particular case where the rotation is made zero \cite{saulo,saulo2}. In this case, it becomes\footnote{We are using natural units, with $8 \pi G = c = 1$.}
\begin{equation}\label{Eq.1}
ds^2=a^2(\eta) \left[d\eta^2-(dx^2+e^{2x}dy^2+dz^2) \right],
\end{equation}
where $\eta$ is the conformal time.

We know that a perfect fluid does not lead to anisotropies. Therefore, an anisotropic scalar field is introduced, which will be responsible for creating a pressure difference in the preferred direction, which we define as being the $z$ direction. In addition, we will introduce a cosmological constant responsible for the present acceleration in the universe expansion. Although the metric is anisotropic, it provides an strictly isotropic CMB (at the background level) due to the existence of a conformal Killing vector parallel to the fluid $4$-velocity \cite{korotkii,obukhov}. In what concerns the distance-redshift relations, we will consider here 288 supernovae Ia with redshifts between 0.0218 and 1.551, collected by the Sloan Digital Sky Survey II (SDSS-II) \cite{Kessler:2009ys} and calibrated with the Multicolor Light Curve Shapes (MLCS2k2) fitter \cite{Jha:2006fm}.

The work is organized as follows. In Section 2 we present the main properties of the RTKO metric, and in Section 3 we build the corresponding cosmological model. In Section 4 we show the method of determining the model parameters from supernovas observations. In Section 5 we obtain the corresponding results for the anisotropic model and we compare it with the $\Lambda$CDM model. In Section 6 we perform a joint analysis which also includes the position of the first acoustic peak in the CMB anisotropy spectrum (which indirectly gives the distance to the last scattering surface) and the current limits to the present value of the Hubble function.  In Section 7 we present our conclusions.

\section{The RTKO metric}

Before proceeding to the construction of the anisotropic model, let us discuss some properties of the RTKO metric.
If we apply the coordinate transformations
\begin{eqnarray}\label{Eq.3,4}
e^{x} & = & \cosh r+\cos \varphi \sinh r,\\
ye^{x} & = & \sin\varphi \sinh r,
\end{eqnarray}
to metric (\ref{Eq.1}), we put it in the cylindrical form
\begin{equation}\label{Eq.5}
ds^2=a^2(\eta)(d\eta^2-dr^2-\sinh^2 rd\varphi^2-dz^2),
\end{equation}
which is invariant under rotations around the $z$-axis. Now, making use of the transformations
\begin{eqnarray}\label{Eq.6,7}
r &=& \chi\sin \theta, \\
z &=& \chi\cos \theta,
\end{eqnarray}
the above line element can be expressed in spherical coordinates as
\begin{equation}\label{Eq.8}
ds^2=a^2(\eta)[d\eta^2-d\chi^2-\chi^2d\theta^2-\sinh^2 (\chi\sin \theta)d\varphi^2].
\end{equation}

The anisotropy is clear in the last term. It will lead to an angular dependence in the angular-diameter distance, as
we shall see.
Interestingly enough, the metric expressed in the last equation reduces to the spatially flat Friedmann-Lema\^{i}tre-Robertson-Walker (FLRW) space-time in the limit of small distances. Indeed, taking the expansion of the hyperbolic sine function up to first-order terms, we obtain
\begin{equation}\label{Eq.9}
ds^2\approx a^2(\eta)[d\eta^2-d\chi^2-\chi^2(d\theta^2+\sin^2 \theta d\varphi^2)],
\end{equation}
i.e., the flat FLRW metric expressed in spherical coordinates. As a consequence, the anisotropic model to be built
will reduce to the standard model in the limit of small distances.

The RTKO metric has Killing vectors
\begin{equation}\label{Eq.10}
\xi _{(1)}=\partial_x-y\partial_y, \;\; \xi_{(2)} = \partial _y, \;\; \xi_{(3)} = \partial _z,
\end{equation}
which classify it as a spatially homogeneous Bianchi type-III metric. It also has a conformal Killing vector
\begin{equation}\label{Eq.11}
\xi_{conf}^{\mu}=\delta_{0}^{\mu}.
\end{equation}
The 4-velocity of a comoving fluid in this space-time is given by
\begin{equation}\label{Eq.12}
u^{\mu}=\frac{dx^\mu}{ds}=\frac{\delta_{0}^{\mu}}{a(\eta)} =\frac{\xi_{conf}^{\mu}}{a(\eta)}.
\end{equation}

Let us consider now the Einstein field equations
\begin{equation}\label{Eq.13}
R_{\nu}^{\mu}-\frac{1}{2}\delta_{\nu}^{\mu}R=T_{\nu}^{\mu}.
\end{equation}
When we use metric (\ref{Eq.1}), we obtain for the energy-momentum tensor the diagonal components
\begin{eqnarray}
T_{0}^{0}a^4 &=& 3a'^2-a^2, \label{Eq.14} \\
T_{1}^{1}a^4 &=& T_{2}^{2}a^4=2aa''-a'^2, \label{Eq.15} \\
T_{3}^{3}a^2 &=& T_{1}^{1}a^2-1, \label{Eq.16}
\end{eqnarray}
while all the non-diagonal ones are zero. Here, the prime denotes derivative with respect to the conformal time. In the
next section we discuss the cosmological models arising from the above metric.

\section{The Anisotropic Model}

Let us define
\begin{equation}\label{Eq.18}
T_{0}^{0}= \epsilon, \;\;\; T_{i}^{i}=-p_{i}
\end{equation}
as the total energy density and pressures of the universe content, where the index $i$ takes the values $1$, $2$ and $3$, representing the pressures in the directions $x$, $y$ e $z$, respectively. Equations (\ref{Eq.14})-(\ref{Eq.16}) then become
\begin{eqnarray}\label{Eq.19,20,21}
\epsilon a^4 &=& 3a'^2-a^2,  \\
p_{1}a^4 &=& p_{2}a^4 = a'^2-2aa'', \\
p_{3}a^2 &=& p_{1}a^2+1.
\end{eqnarray}

This set of equations shows clearly the metric anisotropy, since the pressure along the $z$-direction differs from the others two by an additional term $1/a^2$. The usual components of the universe (radiation, matter and cosmological constant) cannot generate such anisotropy. Let us show that it can be generated, in a self-consistent way, by a massless scalar field minimally coupled to gravity. Such a scalar field satisfies the Klein-Gordon equation
\begin{equation}\label{Eq.22}
\phi_{;\mu\nu}=\frac{1}{\sqrt{-g}}(\sqrt{-g}\Phi_{,\mu}g^{\mu\nu})_{,\nu}=0,
\end{equation}
and has an energy-momentum tensor given by
\begin{equation}\label{Eq.23}
 T_{\nu}^{\mu}=\Phi_{,\nu}\Phi_{,\gamma}g^{\gamma\mu}-\frac{1}{2}\Phi_{,\gamma}\Phi_{,\lambda}g^{\gamma\lambda}\delta_{\nu}^{\mu}.
\end{equation}
Here $g$ is the determinant of the metric tensor $g_{\mu\nu}$, and $\delta_{\nu}^{\mu}$ is the delta of Kronecker. The commas and semicolons refer, respectively, to ordinary and covariant derivatives.

Our scalar field will be given by
\begin{equation}\label{Eq.24}
\Phi(z)=Cz,
\end{equation}
where $C$ is a constant. It is easy to verify that it is solution of equation (\ref{Eq.22}). On the other hand, the non-zero components of its energy-momentum tensor (the diagonal ones) are
\begin{equation}\label{Eq.25}
-T_{0}^{0}=-T_{1}^{1}=-T_{2}^{2}=T_{3}^{3}=-\frac{C^2}{2a^2},
\end{equation}
which lead, through (\ref{Eq.18}), to
\begin{equation}\label{Eq.26}
\epsilon^{(s)}=\frac{C^2}{2a^2},
\end{equation}
\begin{equation}\label{Eq.27}
p_{1}^{(s)}= p_{2}^{(s)}=-p_{3}^{(s)}=-\frac{C^2}{2a^2},
\end{equation}
where the upper index $(s)$ refers to the contribution of the scalar field. We note that the pressures are anisotropic, and the energy density falls with the square of the scale factor. Using this fact we can work out the anisotropy present in Einstein's equations.

Let us define
\begin{equation}\label{Eq.28}
\epsilon\equiv\overline{\epsilon}+\epsilon^{(s)}, \;\;
p_i\equiv \overline{p}_i+p_i^{(s)},
\end{equation}
where the bar refers to the isotropic content (radiation, matter and cosmological constant). Equation (\ref{Eq.16}) takes the
form
\begin{equation}\label{Eq.29}
\overline{p}_3a^2=\overline{p}_1a^2-C^2+1.
\end{equation}
We then see that the anisotropy of the model is absorbed by the scalar field if (and only if) $C^2 = 1$. In this case we
have $\overline{p}_1 = \overline{p}_2 = \overline{p}_3 = \overline{p}$, and the scalar field is simply given by
\begin{equation}\label{Eq.30}
    \Phi(z)=\pm z.
 \end{equation}

The energy density and pressures are now given by
\begin{equation}\label{Eq.31}
\epsilon^{(s)}=\frac{1}{2a^2},
\end{equation}
\begin{equation}\label{Eq.32}
p_{1}^{(s)} = p_{2}^{(s)}=-p_{3}^{(s)}=-\frac{1}{2a^2},
\end{equation}
and the Einstein equations can be rewritten as
\begin{equation}\label{Eq.33}
\overline{\epsilon}a^4 = 3a'^2-\frac{3}{2}a^2,
\end{equation}
\begin{equation}\label{Eq.34}
\overline{p}a^4 =a'^2-2aa''+\frac{a^2}{2}.
\end{equation}

Notably, these equations are precisely those of an open FLRW model with curvature $k = -1/2$. In fact, from (\ref{Eq.33}) we obtain the Friedmann equation
\begin{equation}\label{Eq.35}
    H^2= \frac{\overline{\epsilon}}{3}+\frac{1}{2a^2},
\end{equation}
where we have defined the Hubble parameter $H = a'/a^2$.

On the other hand, from the continuity equation
\begin{equation}\label{Eq.36}
    \frac{\epsilon'}{a}+3H(\epsilon+p)=0
\end{equation}
we can write the energy densities of radiation, non-relativistic matter and the cosmological constant respectively as $\epsilon_r = \frac{A}{a^4}$, $\epsilon_m = \frac{B}{a^3}$,
and $\epsilon_{\Lambda}= \Lambda$,
with $A$, $B$ and $\Lambda$ constants. The total energy density of the isotropic content is then
\begin{equation}\label{Eq.40}
    \overline{\epsilon}=\epsilon_r+\epsilon_m+\epsilon_{\Lambda}=\frac{A}{a^4}+\frac{B}{a^3}+\Lambda.
\end{equation}

The Friedmann equation (\ref{Eq.35}) assumes now the form
\begin{equation}\label{Eq.41}
    H^2=\frac{A}{3a^4}+\frac{B}{3a^3}+\frac{\Lambda}{3}+\frac{1}{2a^2}.
\end{equation}
It can also be written as
\begin{equation}\label{Eq.42}
\Omega^{(s)}=1-\Omega_r-\Omega_m-\Omega_{\Lambda},
\end{equation}
where, for matter, radiation and the cosmological constant, we have defined $\Omega_j = \epsilon_j/3H^2$ ($j$ = $r$, $m$ or $\Lambda$), and
\begin{equation}\label{Eq.43}
    \Omega^{(s)}\equiv\frac{1}{2H^2a^2}.
\end{equation}
Owing to its particular dependence on the scale factor, we will call this last term the ``curvature" of the anisotropic model. However, let us point out that it contains, besides the space-time curvature itself, the contribution of the energy density of the scalar field responsible for the anisotropy.

By using the relation
\begin{equation}\label{Eq.44}
    1+z=\frac{a_0}{a},
\end{equation}
we obtain, from (\ref{Eq.41}) and (\ref{Eq.42}),
\begin{equation}\label{Eq.45}
    \frac{H(z)}{H_0} \equiv E(z)=\sqrt{\Omega_{r0}(1+z)^4+\Omega_{m0}(1+z)^3+(1-\Omega_{r0}-\Omega_{m0}-\Omega_{\Lambda})(1+z)^2+\Omega_{\Lambda}},
\end{equation}
where the index $0$ refers to the present time. On the other hand, the deceleration parameter is defined by
\begin{equation}\label{Eq.46}
    q \equiv 1-\frac{a''a}{a'^2},
\end{equation}
or, as a function of the redshift, by
\begin{equation}\label{Eq.47}
    q(z)=-1+\frac{(1+z)}{H(z)}\frac{dH}{dz},
\end{equation}
while the age parameter is given by
\begin{equation}\label{Eq.48}
    H_0t_0=\int_{0}^{\infty}\frac{1}{1+z}\frac{H_0}{H(z)}dz.
\end{equation}

\section{Parameters estimation}

In order to establish a comparison between the anisotropic model and the standard one (here denominated $k$-$\Lambda$CDM to include the spatially curved cases), we have performed an analysis of $288$ supernovas Ia compiled by the SDSS supernova survey with the fitter MLCS2k2. The best-fit values of the free parameters of both models were determined by means of the usual $\chi^2$ statistics.

\subsection{The $k$-$\Lambda$CDM case}

In this case we use the modulus distance
\begin{equation}\label{Eq.49}
    \mu(z)=\left\{ \begin{array}{ll}
        5\log \left [ {\displaystyle\frac{1+z}{\sqrt{\left | \Omega_{k0}\  \right |}}\sin\left ( \sqrt{ | \Omega_{k0} |  }\int_{0}^{z} \frac{dz}{E(z)} \right)} \right] -5\log(h)+42.38 & (k=+1) \\
\;&\;\\
        5\log \left [ (1+z)\displaystyle\int_{0}^{z} \displaystyle\frac{dz}{E(z)}  \right] -5\log(h)+42.38 & (k=0) \\
\;&\;\\
        5\log \left [ {\displaystyle\frac{1+z}{\sqrt{\left | \Omega_{k0}\  \right |}}\sinh\left ( \sqrt{ |\Omega_{k0} | }\int_{0}^{z} \frac{dz}{E(z)} \right)} \right] -5\log(h)+42.38 & (k=-1),\\
    \end{array}
\right.
\end{equation}
where $h \equiv H_0/100$ km/s-Mpc, and
\begin{equation}\label{Eq.50}
    E(z)=\sqrt{\Omega_{m0}(1+z)^3+(1-\Omega_{r0}-\Omega_{m0}-\Omega_{\Lambda})(1+z)^2+\Omega_{\Lambda}}.
\end{equation}
After marginalising $h$ and fixing $\Omega_{r0}$ with the CMB constraints, the free parameters are $\Omega_{m0}$ and $\Omega_{\Lambda}$.

\subsection{The anisotropic case}

The FLRW metric is isotropic. This means that, given two galaxies with the same redshift, they are at the same luminosity distance from us. However in the case of an anisotropic metric that is not true. For a given redshift we can obtain different distances, i.e., the brightness depends on the distance and on the angle of observation. Therefore, we need to obtain the correct luminosity distance for our metric.

With this goal in mind, we define the angular-diameter distance by \cite{obukhov,koivisto}
\begin{equation}\label{Eq.51}
    {d_{A}}^2=\frac{dA_p}{d\Omega},
\end{equation}
where $dA_p$ is the proper-area element formed by the solid angle $d\Omega  = \sin \theta d\theta d\varphi$. The reason for this choice is the fact that, in an anisotropic metric, the usual definition
\begin{equation}\label{Eq.52}
    d_A=\frac{dl}{d\alpha}
\end{equation}
(where $d\alpha$ is the angle associated to the distance $dl$ between two observed points) would lead to different distances $dl$ for the same $d_A$ as the observation angle $\theta$ changes.

The area element in the anisotropic case can be obtained by taking our spatial metric and fixing $\chi$, which leads to the line element
\begin{equation}\label{Eq.53}
    dl^2=a^2(\eta)[\chi^2d\theta^2+\sinh^2(\chi\sin\theta)d\varphi^2],
\end{equation}
from which we can see that the proper-area element is given by
\begin{equation}\label{Eq.55}
    dA_p(\eta)=a^2(\eta)\chi\sinh(\chi\sin\theta)d\theta d\varphi,
\end{equation}
which, after dividing by $d\Omega$, leads to
\begin{equation}\label{Eq.56}
    d_A(\eta)=a(\eta)\chi \left[ \frac{\sinh(\chi\sin\theta)}{\chi\sin\theta} \right] ^{1/2}.
\end{equation}

In order to perform an observational analysis, we can rewrite equation (\ref{Eq.56}) in terms of the redshift,
\begin{equation}\label{Eq.57}
    d_A(z)=\frac{a_0\chi}{1+z} \left[ \frac{\sinh(\chi\sin\theta)}{\chi\sin\theta} \right] ^{1/2},\;\;\;\; \theta \in (0,\pi).
\end{equation}
On the other hand, the coordinate $\chi$ can be obtained by taking a null geodesic,
\begin{equation}\label{Eq.58}
    d\chi=d\eta\Rightarrow\chi=\frac{1}{a_0H_0}Z(z),
\end{equation}
where we have defined
\begin{equation}\label{Eq.59}
    Z(z)\equiv\int_{0}^{z}\frac{dz'}{E(z')}.
\end{equation}
Therefore, the luminosity distance will be given by
\begin{equation}\label{Eq.61}
    d_L(z)=(1+z)^2 d_A(z)=\frac{(1+z)Z(z)}{H_0} \left[ \frac{\sinh(\sqrt{|\Omega_{k0}|}Z(z)\sin\theta)}{\sqrt{|\Omega_{k0}|}Z(z)\sin\theta} \right] ^{1/2},\;\;\;\; \theta \in (0,\pi).
\end{equation}
We note that, in the limit $\theta \rightarrow 0$ or $\chi \ll 1$, the luminosity distance reduces to that of the spatially flat FLRW model. Let's keep in mind that $|\Omega_{k0}|=2\Omega^{(s)}_0$ (see (\ref{Eq.43})), which permits to rewrite the above equation in terms of $\Omega^{(s)}_0$.

In order to fix the free parameters of the present model, which are $\left\{ \Omega_{m0}, \Omega^{(s)}_0 \right\}$ (after the marginalization of $h$), we use the angular average of the luminosity distance in the interval $[0,\pi]$, for each value of $z$. The average modulus-distance will be given by
\begin{equation}\label{Eq.62}
    \bar{\mu}(z)=5\log[H_0\bar{d}_L(z)]-5\log(h)+42.38,
\end{equation}
where $\bar{d}_L(z)$ is the average of (\ref{Eq.61}) in the interval $\theta \in [0,\pi]$.

\section{Results and discussion}

With the procedure described above, we have obtained the best fit for the $k$-$\Lambda$CDM model with the parameters $\Omega_{m0} = 0.39$, $\Omega_{\Lambda} = 0.58$ and $\Omega_{k0} = 0.03$, with $\chi^2_r = 0.84$. Therefore, in the absence of other priors, the data favors a slightly negative curvature. Within the $2\sigma$ confidence level, we obtain $0.16 < \Omega_{m0} < 0.58$ and $0.12 < \Omega_{\Lambda} < 0.95$. For the curvature we have, within the same confidence level, $-0.51 < \Omega_{k0} < 0.71$.

For the anisotropic case, the best fit corresponds to $\Omega_{m0} = 0.38$, $\Omega_{\Lambda} = 0.58$ and $\Omega^{(s)} = 0.04$, with $\chi^2_r = 0.84$. The $2\sigma$ intervals are $0 < \Omega_{m0} < 0.48$, $0.04 < \Omega_{\Lambda} < 0.67$ and $0 < \Omega^{(s)} < 0.95$. The confidence regions for both models are presented in Fig. 1. In Table I we compare the parameters best values for both models, as well as the present deceleration parameters $q_0$ and the age parameters $H_0 t_0$, obtained from equations (\ref{Eq.47}) and (\ref{Eq.48}), respectively.

In order to set the level of agreement between the anisotropic model and the
observational data, in Fig. 2a we show a comparison between the observed distance moduli and the theoretical predictions. The solid lines correspond to the maximum ($\theta = \pi/2$), average and minimum ($\theta = 0$) values of the distance modulus. In this comparison we have used $\Omega^{(s)} = 0.52$, which is the maximum ``curvature" allowed (within the $2\sigma$ level) when we fix the matter density parameter in $\Omega_{m0} = 0.25$. We can see that, for the range of redshifts available in SDSS, the theoretical predictions are within the observed limits.

\begin{figure}[t]
\centering
{
\includegraphics[height=6cm]{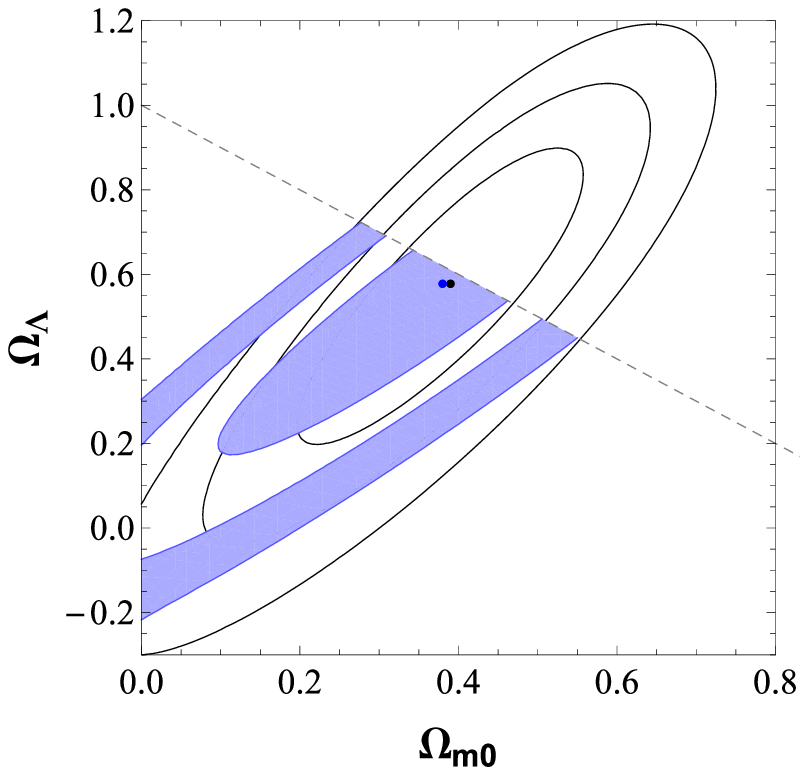}
\label{graf.1a}
}
\qquad
{
\includegraphics[height=6cm]{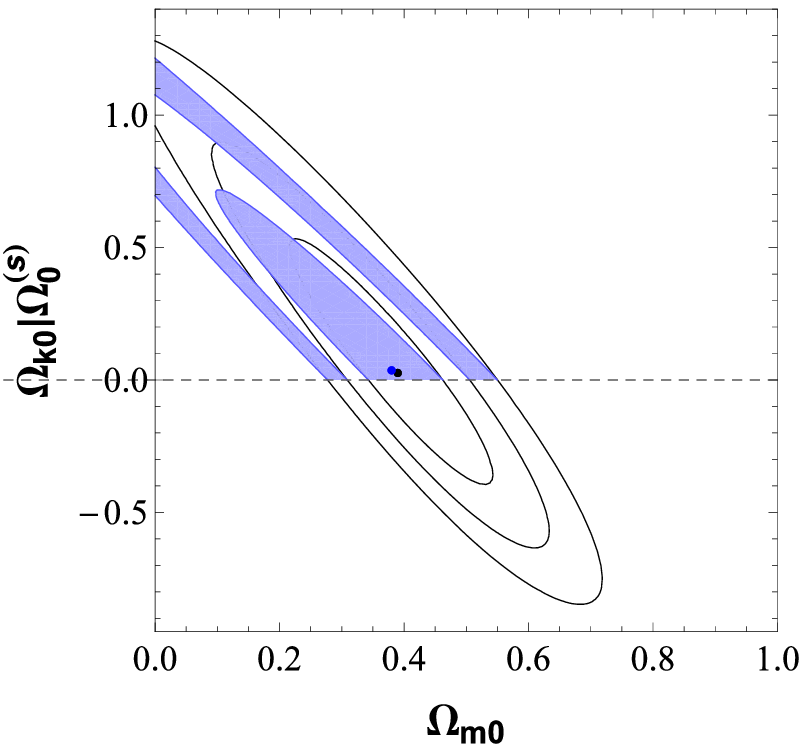}
\label{graf.1b}
}
\caption{\small{Confidence levels $1\sigma$, $2\sigma$ and $3\sigma$ for the $k$-$\Lambda$CDM (white) and anisotropic (blue and white) models with SDSS (MLCS2k2).}}
\end{figure}

\begin{table}[b]
\begin{center}
\begin{tabular}{c|c|c|c|c|c|c}
\hline
Model & $\Omega_{m0}$ & $\Omega_{k0}|\Omega_{0}^{(s)}$ & $\Omega_{\Lambda}$ & ${\chi_{r}}^2$ & $q_0$ & $H_0t_0$ \\
\hline\hline
$k$-$\Lambda$CDM & $0.39$ & $0.03$ & $0.58$ & 0.840 & -0.386 & 0.888\\
\hline
Anisotropic & $0.38$ & $0.04$ & $0.58$ & 0.840 & -0.384 & 0.890 \\
\hline
\end{tabular}
\end{center}
\caption{\small{Parameters best values for the $k$-$\Lambda$CDM and anisotropic models fitted with the sample SDSS (MLCS2k2).}}
\end{table}

As shown above, the best fit requires a small curvature (a spatially open universe) in the $k$-$\Lambda$CDM model. There is then space for confusing
a possible anisotropy with an actual space curvature. In order to verify
in which range of redshifts the two models may be confused, we show in Fig. 2b the average distance modulus as a function of $z$ in both models, for the corresponding best-fit values of their parameters. We can note that both agree with the observational data, and that the curves superpose even for redshifts above $z \approx 2$ (see Fig. 3a). However, when we use the maximum curvature allowed (within $2\sigma$ level) for $\Omega_{m0} = 0.25$ (which corresponds to $\Omega^{(s)} = 0.52$ and $\Omega_{k0} = 0.64$), the models predictions start to diverge (see Fig. 3b), indicating that, above the upper limit of the used data, the test is more sensitive to a possible anisotropy, which could, in principle, be detected.

\begin{figure}[t]
\centering
\subfloat[]
{
\includegraphics[height=5.5cm]{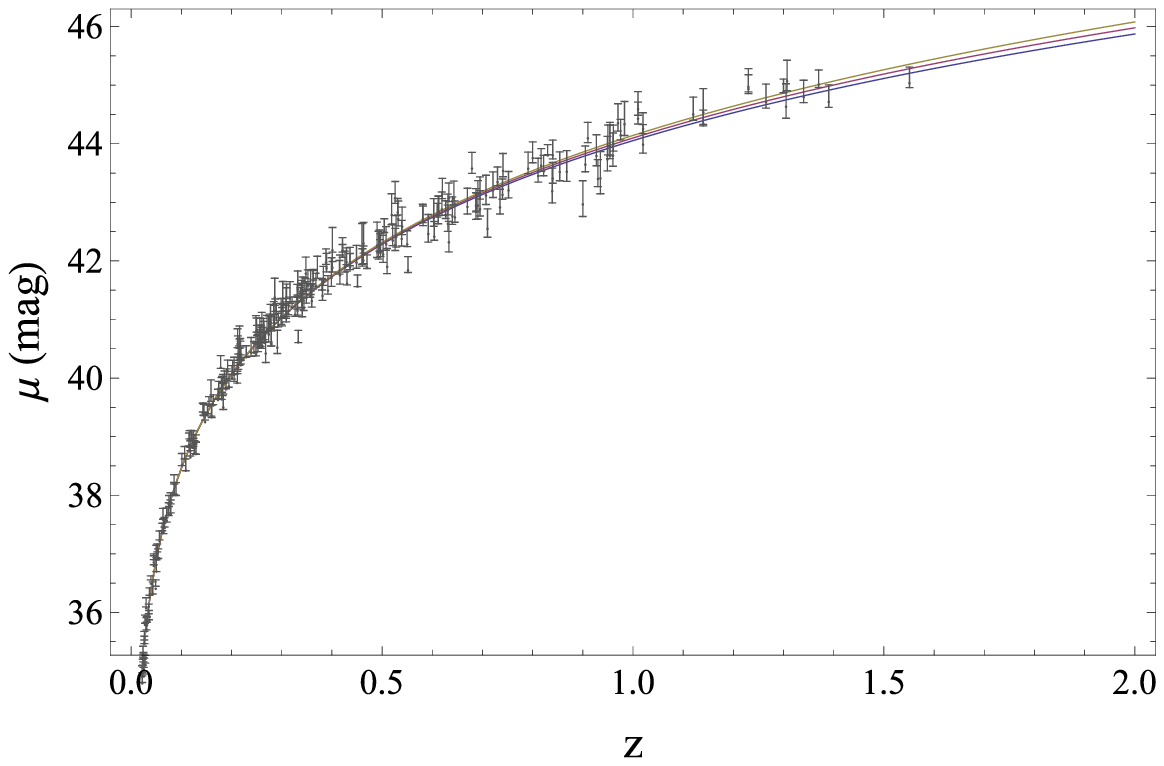}
\label{graf.2}
}
\qquad 
\subfloat[]
{
\includegraphics[height=5.5cm]{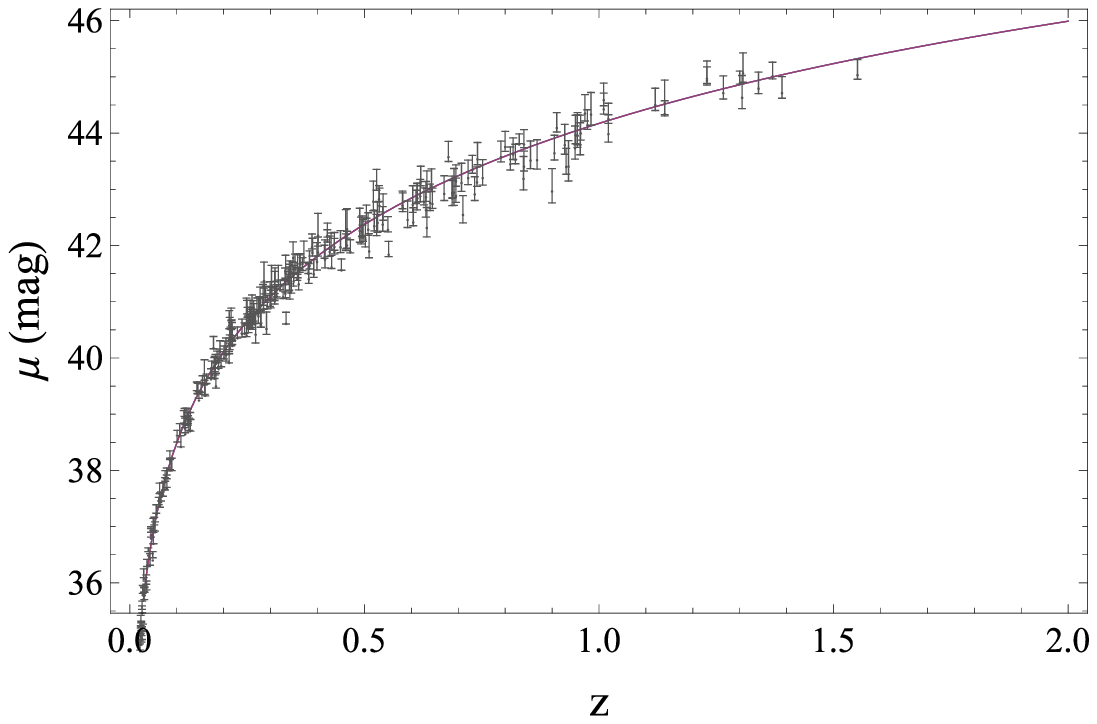}
\label{graf.3}
}
\caption{\small{Distance modulus \emph{versus} redshift: (a) for the anisotropic model with minimum, medium and maximum anisotropies; (b) for the best-fit anisotropic and $k$-$\Lambda$CDM models.}}
\end{figure}

\begin{figure}[th]
\centering
\subfloat[]
{
\includegraphics[height=5.5cm]{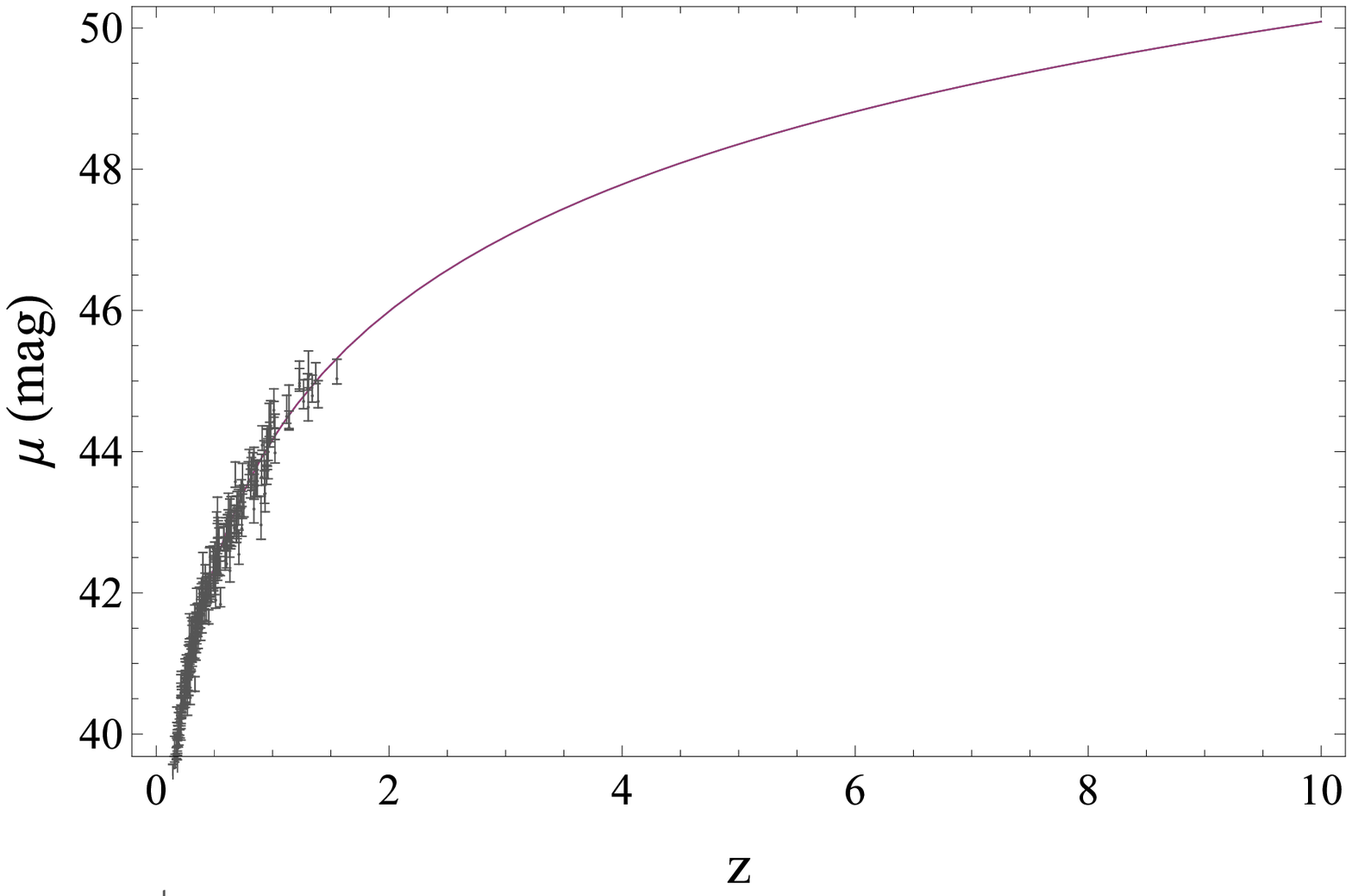}
\label{graf.4}
}
\qquad
\subfloat[]
{
\includegraphics[height=5.5cm]{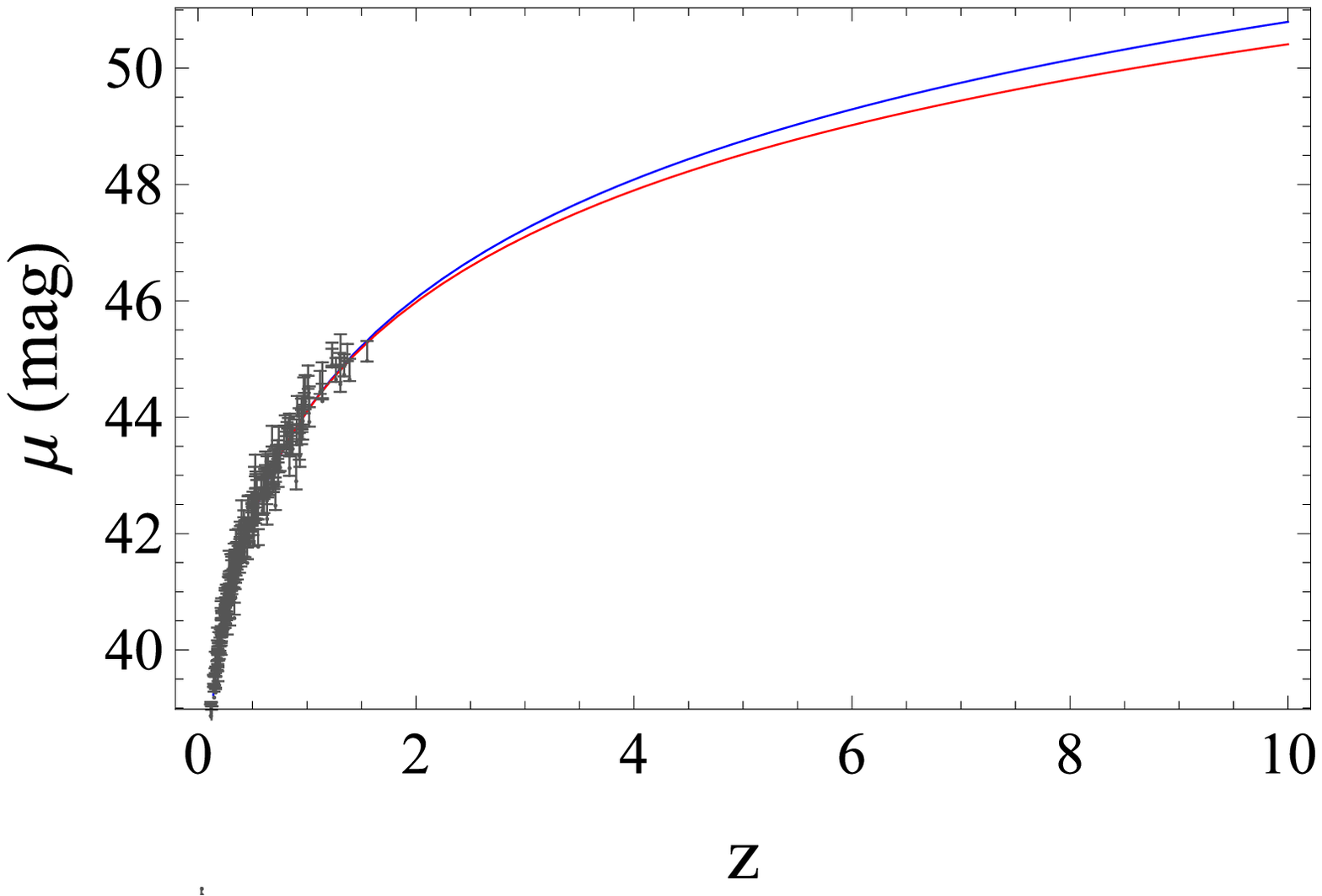}
\label{graf.5}
}
\caption{\small{Distance modulus \emph{versus} redshift: (a) for the best-fit anisotropic and $k$-$\Lambda$CDM models; (b) for both models with maximum curvature parameters. The $\Lambda$CDM is represented by the blue line and the anisotropic model by the red line.}}
\end{figure}

\section{The CMB first peak}

At this stage we can verify at which level our previous results may be altered when we also include in the analysis the position of the first acoustic peak in the CMB anisotropy spectrum, since it may also be considered a distance ladder, used to determine the distance to the last scattering surface. With this goal in mind, we make use of the so-called shift parameter. It is useful for any model whose sound horizon at last scattering and the redshift of radiation-matter equality are the same as in the flat standard model \cite{Elgaroy}. That is the case: since the metric anisotropy scales as $a^{-2}$, it is negligible at high redshifts as compared to the matter and radiation densities. However, in our case a suitable re-definition of the angular-diameter distance is needed, in the same way we have re-defined the luminosity distance in the supernovas analysis.

\subsection{The shift parameter}

The shift parameter in isotropic models is defined by \cite{Elgaroy}
\begin{equation} \label{shiftparameter}
R \equiv \frac{\sqrt{\omega_{m0}}}{\sqrt{\omega_{k0}}}\sin_k(\sqrt{\omega_{k0}}y),
\end{equation}
where $\omega_i=\Omega_ih^2$ and
\begin{equation}\label{y}
y \equiv \int_{a_{ls}}^{1} \frac{da}{\sqrt{\omega_{m0}a + \omega_{k0}a^2 + \omega_{\Lambda}a^4}}=\frac{1}{h}\int_{0}^{z_{ls}}\frac{dz}{E(z)},
\end{equation}
with $E(z)$ now including the contribution of radiation. As usual, $z_{ls}$ stands for the redshift of last scattering, and we have used the notation
\begin{equation}\label{sinx}
    \sin_k(x)=\left\{ \begin{array}{ll}
        \sin(x) & (k=+1), \\
\;&\;\\
        x & (k=0), \\
\;&\;\\
        \sinh(x) & (k=-1).\\
    \end{array}
\right.
\end{equation}

One has
$\omega_{m0}/\omega_{k0}=\Omega_{m0}/\Omega_{k0}$ and
$\sqrt{\omega_{k0}}y = \sqrt{\Omega_{k0}}Z(z)$, where we have again defined
\begin{equation}
Z(z)\equiv \int_{0}^{z_{ls}}\frac{dz}{E(z)}.
\end{equation}
Therefore, the shift parameter can be rewritten as
\begin{equation}\label{R/wm}
\frac{R}{\sqrt{\Omega_{m0}}} = \frac{1}{\sqrt{\Omega_{k0}}}\sin_k[\sqrt{\Omega_{k0}}Z(z)].
\end{equation}
On the other hand, the angular-diameter distance is given by
\begin{equation} \label{da}
d_A = \frac{d_L}{(1+z)^2} = \frac{1}{H_0(1+z)\sqrt{\Omega_{k0}}}\sin_k[\sqrt{\Omega_{k0}}Z(z)].
\end{equation}
By comparing with Eq. (\ref{R/wm}), we then have
\begin{equation}\label{da'}
H_0(1+z_{ls}) d_A = \frac{R}{\sqrt{\Omega_{m0}}}.
\end{equation}

For the anisotropic model we also make use of Eq. (\ref{da'}), but now with the angular-diameter distance appropriately re-defined as (see Eq. (\ref{Eq.61}))
\begin{equation}\label{da_anis}
\bar d_A = \frac{Z(z)}{\pi(1+z_{ls})H_0}\int_{0}^{\pi} \left[ \frac{\sinh(\sqrt{\Omega_{k0}}Z(z)\sin\theta)}{\sqrt{\Omega_{k0}}Z(z)\sin\theta} \right]^{1/2}d\theta.
\end{equation}
In this way, we obtain the shift-parameter
\begin{equation}\label{R_anis}
\bar{R} = \frac{\sqrt{\Omega_{m0}}Z(z)}{\pi}\int_{0}^{\pi} \left[ \frac{\sinh(\sqrt{\Omega_{k0}}Z(z)\sin\theta)}{\sqrt{\Omega_{k0}}Z(z)\sin\theta} \right]^{1/2}d\theta.
\end{equation}

\subsection{Joint analysis}

For the jointy analysis we use
\begin{equation}\label{chi2_comb}
\chi^2 = \chi^2_{SN} + \chi^2_{CMB} + \chi^2_{prior},
\end{equation}
 or, more precisely,
\begin{equation}\label{chi2_comb2}
\chi^2 = \sum_{i=1}^{288}\frac{[ \mu_{teo}(z_i|\Omega_{m0},\Omega_{\Lambda},h)-\mu_{obs}(z_i) ]^2}{\sigma_{\mu}^2+\sigma_{z}^2+\sigma_{sist}^2} + \frac{[ \bar{R}(\Omega_{m0},\Omega_{\Lambda},h) -1.710 ]^2}{(0.019)^2} + \frac{(h-0.72)^2}{(0.08)^2}.
\end{equation}

In the above statistics we have also included a prior for the present Hubble parameter, given by $h = 0.72 \pm 0.08$ \cite{Freedman}. For the shift parameter we take $\bar{R} = 1.710 \pm 0.019$ \cite{Komatsu}. For the radiation density we will use the standard-model value $\Omega_{r0} h^2 = 0.00004116$ \cite{Komatsu}.

The best-fit results obtained for the $k$-$\Lambda$CDM are $\Omega_{m0}= 0.40$ and $\Omega_{\Lambda} = 0.63$. Within the $2\sigma$ confidence level we have $0.37<\Omega_{m0}< 0.45$ and $0.60<\Omega_{\Lambda}< 0.65$.  For the anisotropic model the best-fit results are $\Omega_{m0}= 0.32$ and $\Omega_{\Lambda} = 0.67$, with $2\sigma$ intervals $0.29<\Omega_{m0}< 0.37$ and $0.63<\Omega_{\Lambda}< 0.70$. The corresponding confidence regions are given in Figure 4. We can see that the inclusion of CMB in the analysis imposes more restrictive limits to the densities. Nevertheless, we still have freedom for up to $8\%$ of anisotropy at $2\sigma$ level.

\begin{figure}[t]
\centering
{
\includegraphics[height=7cm]{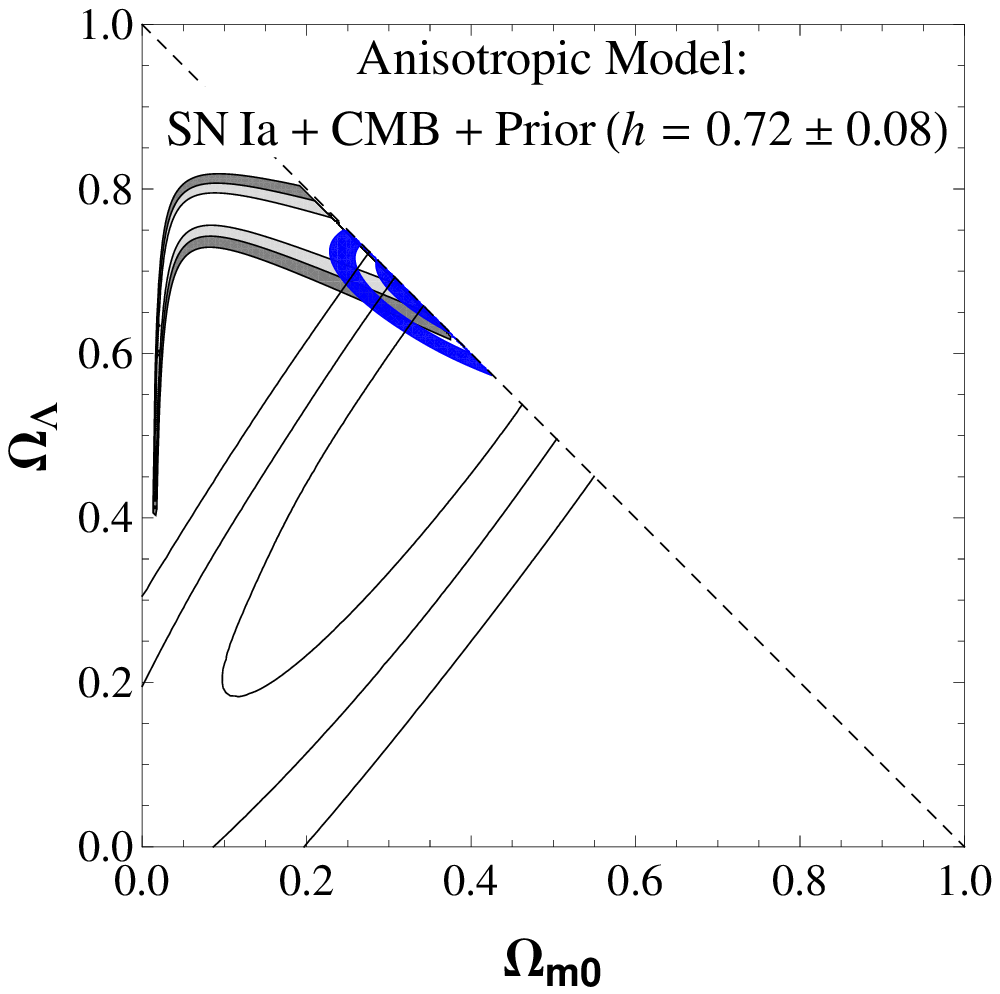}
\label{graf.4a}
}
\qquad
{
\includegraphics[height=7cm]{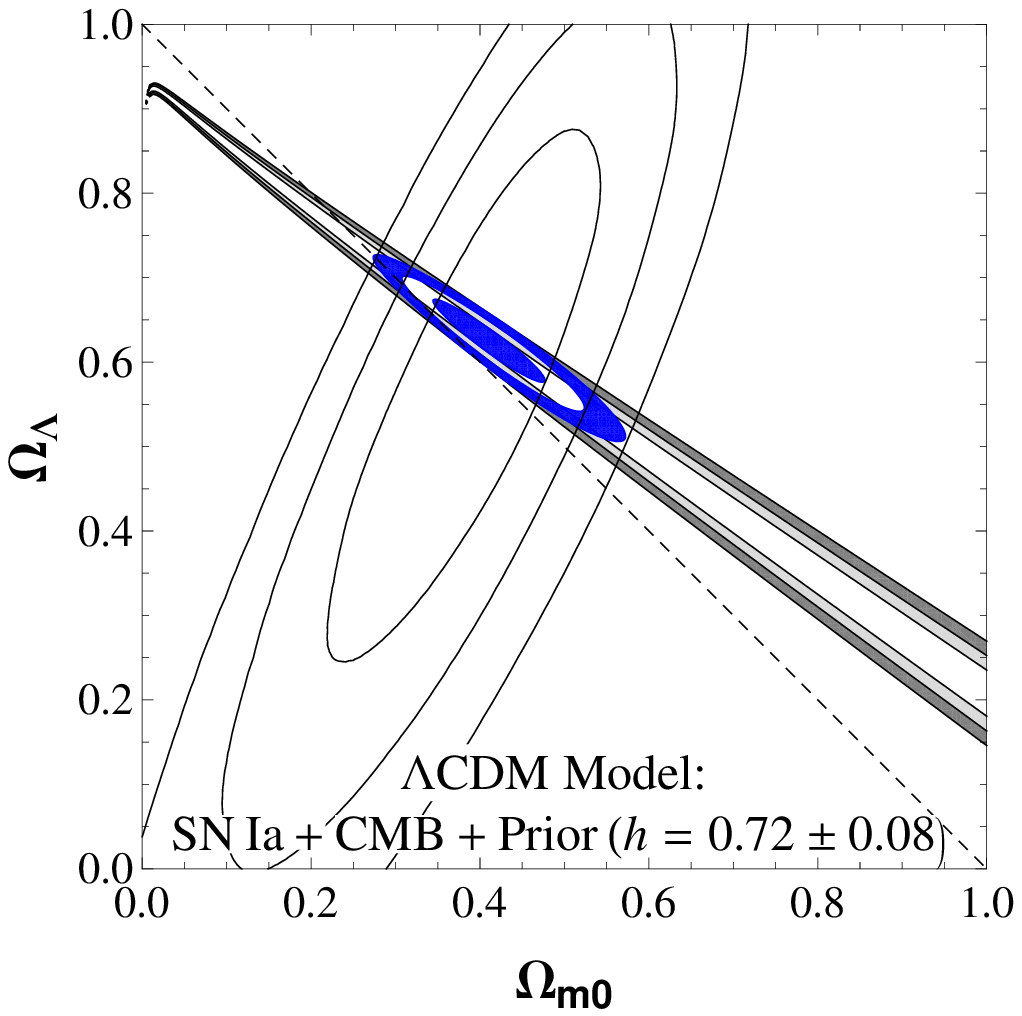}
\label{graf.4b}
}
\caption{\small{Confidence levels (in blue) for the joint analysis SDSS + CMB + $H_0$, for the $k$-$\Lambda$CDM (right) and anisotropic (left) models.}}
\label{}
\end{figure}

\section{Conclusion}

We have studied the observational viability of an anisotropic cosmology based on a Bianchi III metric which generalizes the G\"odel solution, in the particular case where the cosmic rotation is made zero. As shown elsewhere, such metric leads to an exactly isotropic CMB (at the background level) and, in addition, it is an exact solution of the Einstein equations in the presence of an anisotropic, minimally coupled scalar field.

This metric induces an anisotropy in the luminosity distances of galaxies with same redshifts but observed in different directions. This effect can, in principle, be used to detect a preferential axis in the sky through the observation of supernovas Ia at high redshifts. In the present work we have made use of the SDSS supernova sample calibrated with the MLCS2k2 fitter, since it is less dependent on a $\Lambda$CDM fiducial model when compared to SALT or SALT II fitters. For comparison, we have also analysed the $\Lambda$CDM model, with any possible curvature.

Our analysis has shown that, in both cases, a small curvature/anisotropy is allowed by observation, a conclusion not altered when the observed position of the first acoustic peak in the CMB anisotropy spectrum is included in a joint analysis. On the other hand, the present data do not distinguish between the two models, even when we consider the maximum permitted curvature/anisotropy. A more robust conclusion about the existence of an anisotropy at cosmological scales may, however, be reached when supernova data with higher redshifts will be available.

\section*{Acknowledgements}

The authors are thankful to CNPq (Brazil) for the grants under which this work was carried out, and to Rita Novaes for the manuscript reading.

{}

\end{document}